\documentclass[12pt]{iopart}
\usepackage{epsf}
\begin{document}
\title{\Large Stub model for dephasing in a quantum dot}
\author{C. W. J. Beenakker and B. Michaelis}
\address{Instituut-Lorentz, Universiteit Leiden,\\
P.O. Box 9506, 2300 RA Leiden, The Netherlands}
\begin{abstract}
As an alternative to B\"{u}ttiker's dephasing lead model, we examine a dephasing {\em stub}. Both models are phenomenological ways to introduce decoherence in chaotic scattering by a quantum dot. The difference is that the dephasing lead opens up the quantum dot by connecting it to an electron reservoir, while the dephasing stub is closed at one end. Voltage fluctuations in the stub take over the dephasing role from the reservoir. Because the quantum dot with dephasing lead is an open system, only {\em expectation values\/} of the current can be forced to vanish at low frequencies, while the outcome of an individual measurement is not so constrained. The quantum dot with dephasing stub, in contrast, remains a closed system with a vanishing low-frequency current at each and every measurement. This difference is a crucial one in the context of quantum algorithms, which are based on the outcome of individual measurements rather than on expectation values. We demonstrate that the dephasing stub model has a parameter range in which the voltage fluctuations are sufficiently strong to suppress quantum interference effects, while still being sufficiently weak that classical current fluctuations can be neglected relative to the nonequilibrium shot noise.
\end{abstract}
\submitto{Journal of Physics A, special issue on ``Trends in Quantum Chaotic Scattering''}

\section{Introduction}

The dephasing lead model was introduced by B\"{u}ttiker in 1986 as a phenomenological description of the loss of coherence in quantum electron transport \cite{But86}. A microscopic theory of dephasing by electron-electron interactions exists in disordered systems \cite{Alt85,Ale99}, but not in (open) chaotic systems. For that reason, experiments on conduction through a chaotic quantum dot are routinely modeled by B\"{u}ttiker's device --- with considerable success \cite{Mar92,Cla95,Hui98,Hui99}.

\begin{figure}
\epsfbox{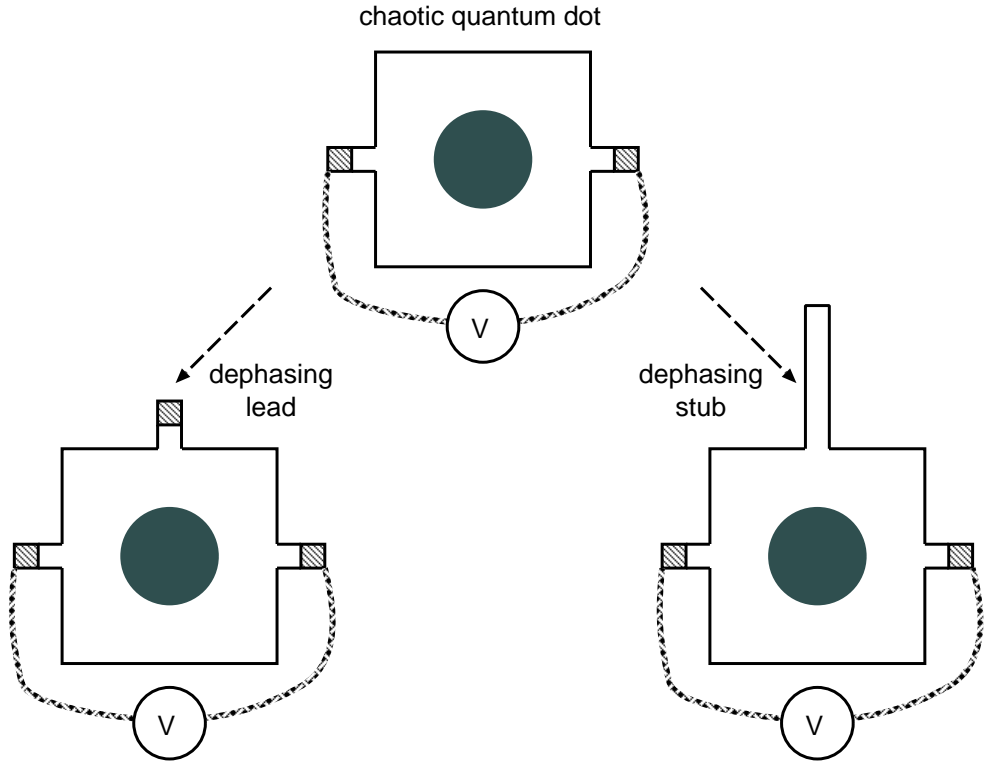}
\caption{
Illustration of two phenomenological ways to model dephasing in a quantum dot. The top panel shows the fully phase coherent system, while the two lower panels introduce dephasing either by means of a dephasing lead (left), or by means of a dephasing stub (right). The shaded rectangles indicate electron reservoirs and the encircled $V$ indicates a voltage source. The voltage on the electron reservoir connected to the dephasing lead is adjusted such that it draws no current when averaged over many measurements. The dephasing stub, in contrast, draws no current at each measurement.
\label{figstublead}}
\end{figure}

An alternative phenomenogical approach, introduced by Vavilov and Aleiner in 1999, is to introduce dephasing by means of a fluctuating time-dependent electric field \cite{Vav99}. This approach was reformulated as the dephasing stub model by Polianski and Brouwer \cite{Pol03}. The two models, dephasing lead and dephasing stub, are illustrated in Fig.\ \ref{figstublead}. Polianski and Brouwer showed that the weak localization correction to the conductance is suppressed in the same way by dephasing in the two models. 

The key difference between the dephasing lead and the dephasing stub is that the former system is open while the latter system is closed. Because the quantum dot is connected to an electron reservoir by the dephasing lead, only {\em expectation values\/} of the current can be forced to vanish at low frequencies; the outcome of an individual measurement is not so constrained. The quantum dot with the dephasing stub remains a closed system with a vanishing low-frequency current at each and every measurement. The difference is irrelevant for the time-averaged current (and therefore for the conductance), but not for the time-dependent current fluctuations. Indeed, recent studies of shot noise find differences between the two models of dephasing \cite{Mar04a,Mar04b,Cle04}.

In the context of quantum information processing, the dephasing stub model seems a more natural starting point than the dephasing lead model. This is because quantum algorithms are based on the outcome of individual measurements rather than on expectation values, so the model for dephasing should conserve the particle number at each measurement --- rather than only on average.

The existing dephasing stub model, however, has an undesired feature that prevents its use as a phenomenological model for dephasing. Ref.\ \cite{Pol03} considers a {\em short\/} dephasing stub, in which the mean dwell time of an electron is negligibly small compared to the mean dwell time in the quantum dot. The voltage fluctuations in a short stub drive the quantum dot out of equilibrium, as is manifested by a nonzero noise power at zero temperature and zero applied voltage \cite{Pol03,Pol02}. We need to avoid this, since true dephasing should have no effect in equilibrium. The original dephasing lead model had this property, that it preserved equilibrium. In this paper we will remove this undesired feature of the dephasing stub model, by demonstrating that a {\em long\/} dephasing stub can be an effective dephaser without driving the quantum dot appreciably out of equilibrium. It therefore combines the two attractive features of the existing models for dephasing: (1) Current conservation for individual measurements and (2) preservation of equilibrium.

\section{Formulation of the problem}

The characteristic properties of quantum dot and stub are their level spacings $\delta_{\rm dot},\delta_{\rm stub}$ and the contact conductances $g_{\rm dot},g_{\rm stub}$ (in units of the conductance quantum $e^{2}/h$, ignoring spin). We assume that the dot is coupled to electron reservoirs by ballistic point contacts, with $g_{\rm dot}=N_{\rm dot}$ the total number of channels in these point contacts. The coupling between dot and stub is via a tunnel barrier with conductance $g_{\rm stub}=N_{\rm stub}\Gamma$ (where $N_{\rm stub}$ is the number of channels and $\Gamma$ is the transmission probability per channel). The limit $N_{\rm stub}\rightarrow\infty$, $\Gamma\rightarrow 0$ at fixed $g_{\rm stub}$ ensures spatial uniformity of the dephasing \cite{Bro97}.

We assume that the dynamics in the quantum dot and in the stub is chaotic. We define the Heisenberg times $\tau_{\rm H,dot}=h/\delta_{\rm dot}$, $\tau_{\rm H,stub}=h/\delta_{\rm stub}$ and the dwell times $\tau_{\rm D,dot}=\tau_{\rm H,dot}/g_{\rm dot}$, $\tau_{\rm D,stub}=\tau_{\rm H,stub}/g_{\rm stub}$. The dwell time $\tau_{\rm D,dot}$ refers to the original quantum dot, before it was coupled to the stub.

In the short-stub model of Polianski and Brouwer \cite{Pol03} the scattering by the stub is time dependent but instantaneous, described by an $N_{\rm stub}\times N_{\rm stub}$ scattering matrix $R(t)$ that depends on a single time argument only. We wish to introduce a delay time in the stub, so we need a scattering matrix $R(t',t)$ that depends on an initial time $t$ and a final time $t'$. The difference $t'-t> 0$ is the time delay introduced by the stub. The reflection by the tunnel barrier is incorporated in $R$, so that it also contains an instantaneous contribution $\delta(t-t')(1-\Gamma)$ times the unit matrix.

The voltage fluctuations are introduced by a spatially random potential $V_{\rm stub}({\bf r},t)$ of the stub, with Gaussian statistics characterized by a mean $v(t)$ and standard deviation $\sigma(t)$. Averages $\langle\cdots\rangle$ over sample-to-sample fluctuations are taken using the methods of random-matrix theory \cite{Bee97}, in the metallic regime $g_{\rm stub}\gg 1$, $g_{\rm dot}\gg 1$.

\section{Diffuson and cooperon}

Quantum corrections to transport properties in the metallic regime are described by two propagators, the diffuson and the cooperon, each of which is determined by an integral equation (the Dyson equation). In disordered systems, the Dyson equation results from an average over random impurity configurations \cite{Lee85}. In the ensemble of chaotic quantum dots, it results from an average over the circular ensemble of scattering matrices \cite{Pol03}. 

The Dyson equation for the diffuson $D$ has the form
\begin{eqnarray}
\fl\tau_{\rm D,dot}D(t,t-\tau;s,s-\tau)=\theta(\tau)e^{-\tau/\tau_{0}}+\theta(\tau)\int_{0}^{\tau}d\tau_{1}\,D(t,t-\tau_{1};s,s-\tau_{1})\nonumber\\
\fl\;\;\;\mbox{}\times\int_{0}^{\tau-\tau_{1}}d\tau_{2}\,e^{-(\tau-\tau_{1}-\tau_{2})/\tau_{0}}N_{\rm stub}D_{\rm stub}(t-\tau_{1},t-\tau_{1}-\tau_{2};s-\tau_{1},s-\tau_{1}-\tau_{2}),\nonumber\\
\label{diffusonDyson}
\end{eqnarray}
where the kernel $D_{\rm stub}$ is the diffuson of the stub,
\begin{equation}
\langle{\rm tr}\,R(t,t-\tau)R^{\dagger}(s,s-\tau')\rangle=\delta(\tau-\tau')N_{\rm stub}D_{\rm stub}(t,t-\tau;s,s-\tau),\label{diffusonstubnodephasing}
\end{equation}
and we have defined $\tau_{0}=\tau_{\rm H,dot}/(N_{\rm dot}+N_{\rm stub})$. Eq.\ (\ref{diffusonDyson}) reduces to the Dyson equation of Ref.\ \cite{Pol03} if the time delay in the stub is disregarded ($\tau_{2}\rightarrow 0$).

In the presence of time-reversal symmetry we also need to consider the cooperon, determined by the Dyson equation
\begin{eqnarray}
\fl\tau_{\rm D,dot}C(t,t-\tau;s+\tau,s)=\theta(\tau)e^{-\tau/\tau_{0}}+\theta(\tau)\int_{0}^{\tau}d\tau_{1}\,C(t,t-\tau_{1};s+\tau_{1},s)\nonumber\\
\fl\;\;\;\mbox{}\times\int_{0}^{\tau-\tau_{1}}d\tau_{2}\,e^{-(\tau-\tau_{1}-\tau_{2})/\tau_{0}}
N_{\rm stub}C_{\rm stub}(t-\tau_{1},t-\tau_{1}-\tau_{2};s+\tau_{1}+\tau_{2},s+\tau_{1}),\label{cooperonDyson}
\end{eqnarray}
where the cooperon of the stub is defined by
\begin{equation}
\langle{\rm tr}\,R(t,t-\tau)R^{\ast}(s+\tau',s)\rangle=\delta(\tau-\tau')N_{\rm stub}C_{\rm stub}(t,t-\tau;s+\tau,s).\label{cooperonstubnodephasing}
\end{equation}

\subsection{Without voltage fluctuations}

Let us first consider the case of a stub without voltage fluctuations. Then only the time difference $\tau$ plays a role, and not the actual times $t,s$. We abbreviate $D(t,t-\tau;s,s-\tau)\equiv D(\tau)$. The cooperon need not be calculated separately, because $C(t,t-\tau;s+\tau,s)=D(\tau)$.

The diffuson of the stub is
\begin{equation}
D_{\rm stub}(\tau)=(1-\Gamma)\delta(\tau)+\Gamma\tau_{\rm D,stub}^{-1}\theta(\tau)e^{-\tau/\tau_{\rm D,stub}}.
\end{equation}
Substitution in the Dyson equation (\ref{diffusonDyson}) gives
\begin{eqnarray}
\fl\tau_{\rm D,dot}D(\tau)=\theta(\tau)e^{-\tau/\tau_{0}}+\theta(\tau)\int_{0}^{\tau}d\tau'\,D(\tau')\left[a e^{-(\tau-\tau')/\tau_{0}}+b e^{-(\tau-\tau')/\tau_{\rm D,stub}}\right],\\
a=N_{\rm stub}+\frac{N_{\rm stub}\Gamma\tau_{\rm D,stub}}{\tau_{0}-\tau_{\rm D,stub}},\;\;b=\frac{N_{\rm stub}\Gamma\tau_{0}}{\tau_{\rm D,stub}-\tau_{0}}.
\end{eqnarray}

This integral equation can be solved by Fourier transformation, or alternatively, by substituting the Ansatz $D(\tau)=\theta(\tau)(\alpha e^{-x\tau}+\beta e^{-y\tau})$ and solving for the coefficients $\alpha,\beta,x,y$. The result is
\begin{eqnarray}
\fl\tau_{\rm D,dot}D(\tau)=\theta(\tau)\frac{x_{+}-1/\tau_{\rm D,stub}}{x_{+}-x_{-}}e^{-x_{+}\tau}+ \theta(\tau)\frac{x_{-}-1/\tau_{\rm D,stub}}{x_{-}-x_{+}}e^{-x_{-}\tau},\label{Dnodephasinga}\\
\fl x_{\pm}=\frac{1}{2}\left(\frac{1}{\tau_{\rm D,stub}}+\frac{1}{\tau_{\rm D,dot}}+\frac{1}{\tau_{\phi}}\right)\pm\frac{1}{2}\sqrt{\left(\frac{1}{\tau_{\rm D,stub}}-\frac{1}{\tau_{\rm D,dot}}-\frac{1}{\tau_{\phi}}\right)^{2}+\frac{4}{\tau_{\phi}\tau_{\rm D,stub}}}.\label{Dnodephasingb}
\end{eqnarray}
The time $\tau_{\phi}=\tau_{\rm H,dot}/N_{\rm stub}\Gamma$ corresponds to the dephasing time in the dephasing lead model. One can verify that the solution (\ref{Dnodephasinga}) satisfies the unitarity relation
\begin{equation}
\int_{0}^{\infty}D(\tau)\,d\tau=1.\label{sumrule}
\end{equation}

Note that the two parameters $N_{\rm stub}$ and $\Gamma$ always appear together as $N_{\rm stub}\Gamma$. This is a simplying feature of the metallic regime $g_{\rm stub}=N_{\rm stub}\Gamma\gg 1$. Since in this regime the tunnel barrier in the stub only serves to renormalize the number of channels, we might as well have assumed a ballistic coupling of the quantum dot to the stub. To simplify the formulas, we will take $\Gamma=1$ in what follows.

\subsection{With voltage fluctuations}
In the presence of a time dependent potential, the diffuson and cooperon of the stub are given by
\begin{eqnarray}
\fl D_{\rm stub}(t,t-\tau;s,s-\tau)=\tau_{\rm D,stub}^{-1}\theta(\tau)e^{-\tau/\tau_{\rm D,stub}}\exp\left(-i\int^{t}_{t-\tau}d\tau'\,\bigl[v(\tau')-v(s-t+\tau')\bigr]\right)\nonumber\\
\mbox{}\times\exp\left(-2\tau_{\rm H,stub}\int^{t}_{t-\tau}d\tau'\,\bigl[\sigma(\tau')-\sigma(s-t+\tau')\bigr]^{2}\right)\nonumber\\
=C_{\rm stub}(t,t-\tau;s,s-\tau).
\end{eqnarray}
(We have set $\hbar=1$.)

To simplify the solution of the Dyson equation, we assume that the spatial average $v(t)$ of the potential $V_{\rm stub}({\bf r},t)$ in the stub vanishes and that the standard deviation $\sigma(t)$ has Gaussian fluctuations in time with moments
\begin{equation}
\langle\sigma(t)\rangle=0,\;\;
\langle \sigma(t)\sigma(t')\rangle=\frac{\gamma\tau_{c}}{4\tau_{\rm H,stub}}\delta_{\tau_{c}}(t-t').
\end{equation}
The time $\tau_{c}$ is the correlation time of the fluctuating potential [setting the width of the regularized delta function $\delta_{\tau_{c}}(t)$] and the rate $\gamma$ is a measure of its strength. 
The average of $D_{\rm stub}$ over the Gaussian white noise is
\begin{eqnarray}
&&\langle D_{\rm stub}(t,t-\tau;s,s-\tau)\rangle=\theta(\tau)\tau_{\rm D,stub}^{-1}\exp\bigl[-\tau Q(s-t)\bigr],\\
&&Q(s-t)=\left\{\begin{array}{ll}
1/\tau_{\rm D,stub}+\gamma&{\rm if}\;\;|s-t|\gg\tau_{c},\\
1/\tau_{\rm D,stub}&{\rm if}\;\;|s-t|\ll\tau_{c}.\label{Qdef}
\end{array}\right.
\end{eqnarray}

For $\tau_{\rm D,stub}\gg\tau_{c}$ the voltage fluctuations in the stub are self-averaging, which means that we may substitute the kernel $D_{\rm stub}$ in the Dyson equation (\ref{diffusonDyson}) by its average $\langle D_{\rm stub}\rangle$. The solution has the same form as the result (\ref{Dnodephasinga}) without voltage fluctuations, but with different coefficients:
\begin{eqnarray}
\fl\tau_{\rm D,dot}D(t,t-\tau,s,s-\tau)=\theta(\tau)\frac{y_{+}-Q(s-t)}{y_{+}-y_{-}}e^{-y_{+}\tau}+ \theta(\tau)\frac{y_{-}-Q(s-t)}{y_{-}-y_{+}}e^{-y_{-}\tau},\label{Dwithdephasinga}\\
\fl y_{\pm}=\frac{1}{2}\left(Q(s-t)+\frac{1}{\tau_{\rm D,dot}}+\frac{1}{\tau_{\phi}}\right)\pm\frac{1}{2}\sqrt{\left(Q(s-t)-\frac{1}{\tau_{\rm D,dot}}-\frac{1}{\tau_{\phi}}\right)^{2}+\frac{4}{\tau_{\phi}\tau_{\rm D,stub}}}.\label{Dwithdephasingb}
\end{eqnarray}
The cooperon is again given by the same expression, $C(t,t-\tau;s,s-\tau)=D(t,t-\tau;s,s-\tau)$.

\section{Transport properties}

A current is passed through the quantum dot by connecting $N_{\rm dot}/2$ channels to one electron reservoir and $N_{\rm dot}/2$ channels to another reservoir at a higher electrical potential $V_{\rm bias}$. We calculate the conductance and the shot noise power of the quantum dot.

\subsection{Weak localization}

The weak localization correction $\delta G$ to the classical conductance $G_{\rm 0}=N_{\rm dot}/2$ of the quantum dot is given by the time integral of the cooperon \cite{Pol03},
\begin{equation}
\delta G=-\frac{1}{4}\int_{0}^{\infty}d\tau\,C(0,-\tau;\tau,0).\label{deltaGdef}
\end{equation}
As before, the conductance is measured in units of the conductance quantum $e^{2}/h$ (ignoring spin). 

The function $C(0,-\tau;\tau,0)$ is given by Eq.\ (\ref{Dwithdephasinga}) with $Q(s-t)\rightarrow Q(\tau)$. For $\tau_{\rm D,stub}\gg\tau_{c}$ we may substitute $Q=1/\tau_{\rm D,stub}+\gamma$, cf.\ Eq.\ (\ref{Qdef}). Carrying out the integration we obtain the expected algebraic suppression of the weak localization correction due to dephasing \cite{Pol03},
\begin{equation}
\delta G=-{\textstyle\frac{1}{4}}(1+\tau_{\rm D,dot}/\tau^{\ast})^{-1},\;\;
\tau^{\ast}=\tau_{\phi}(1+1/\gamma\tau_{\rm D,stub}).\label{deltagresult}
\end{equation}
For $\gamma\tau_{\rm D,stub}\gg 1$ (strong dephasing in the stub) the dephasing time $\tau^{\ast}$ of the dephasing stub model becomes the same as the dephasing time $\tau_{\phi}$ of the dephasing lead model \cite{Bro97,Bar95,Ale96}.

\subsection{Shot noise}

In the absence of a fluctuating potential, the zero-temperature noise power is given by the shot noise formula \cite{Jal94}
\begin{equation}
S_{\rm shot}={\textstyle\frac{1}{4}}eV_{\rm bias}G_{0}.
\end{equation}
The fluctuating potential drives the quantum dot out of equilibrium, adding a contribution $\Delta S$ to the total noise power $S=S_{\rm shot}+\Delta S$. We would like to minimize this classical contribution, since it is unrelated to dephasing.

The general expression for $\Delta S$ contains a product of two diffusons \cite{Pol03},
\begin{eqnarray}
&&\Delta S=\frac{N_{\rm dot}e^{2}}{2}\int_{0}^{\infty}dt\,\bigl[1-K(t)^{2}\bigr]\frac{1-\sin^{2}(eV_{\rm bias}t/2\hbar)}{2\pi^{2} t^{2}},\\
&&K(t)=\int_{0}^{\infty}d\tau\, D(0,-\tau,t,t-\tau).
\end{eqnarray}
Substitution of Eq.\ (\ref{Dwithdephasinga}) gives
\begin{eqnarray}
K(t)&=&\left[1+\frac{\tau_{\rm D,dot}}{\tau_{\phi}}\left(1-\frac{1}{\tau_{\rm D,stub}Q(t)}\right)\right]^{-1}\nonumber\\
&=&\left\{\begin{array}{cc}
1&{\rm if}\;\;t\ll\tau_{c},\\
(1+\tau_{\rm D,dot}/\tau^{\ast})^{-1}&{\rm if}\;\;t\gg\tau_{c}.
\end{array}\right.
\end{eqnarray}

For strong dephasing ($\tau_{\rm D,dot}/\tau^{\ast}\gg 1$) the noise $\Delta S$ from the fluctuating potential saturates at a value of order $\Delta S_{\rm max}\simeq N_{\rm dot}e^{2}/\tau_{c}$. This is negligibly small relative to $S_{\rm shot}$ for sufficiently large bias voltages $V_{\rm bias}\gg\hbar/e\tau_{\rm c}$. These are still small bias voltages on the scale of the Thouless energy $E_{T}=\hbar/\tau_{\rm D,dot}$, provided that $\tau_{\rm D,dot}\ll\tau_{c}$. Combined with our earlier requirement $\tau_{c}\ll\tau_{\rm D,stub}$ of rapid fluctuations, we conclude that the noise generated by the fluctuating potential can be neglected in the regime
\begin{equation}
\tau_{\rm D,dot}\ll\hbar/V_{\rm bias}\ll\tau_{c}\ll\tau_{\rm D,stub}.
\end{equation}

It is the separation of dwell times $\tau_{\rm D,dot}\ll\tau_{\rm D,stub}$, characteristic of the long-stub model, that makes it possible to enter this regime in which the fluctuating potential can be dephasing without being noisy. In contrast, in the short-stub model of Ref.\ \cite{Pol03} one is in the opposite regime $\tau_{\rm D,stub}\ll\tau_{\rm D,dot}$ in which the voltage fluctuations are either too weak to cause dephasing, or so strong that they dominate over the shot noise.

\section{Conclusion}

A fluctuating time-dependent potential in a conductor has both a quantum mechanical effect (destroying phase coherence) and a classical effect (driving the system out of equilibrium). The former effect shows up in the suppression of weak localization, while the latter effect manifests itself in the noise power.
Both effects have been studied extensively in the literature \cite{Vav99,Pol03,See03,For05}, and both effects are important if one is describing a conductor in a real microwave field. However, if the voltage fluctuations are to serve as a phenomenological model of dephasing, e.g.\ by electron-electron interactions, then one needs to retain only the former effect --- since shot noise should be insensitive to dephasing \cite{Jon96}.

The key question we have addressed in this work, is whether a fluctuating potential can be dephasing without being noisy. We have found that this is indeed possible, provided that the potential fluctuates not in the conductor itself but in a spatially separated and weakly coupled region (the stub). The dephasing stub is a fictitious device, much like the dephasing lead \cite{But86}. We expect that the dephasing stub model will be useful as a phenomenological description of decoherence in problems where one would rather not open up the system to an electron reservoir (as one needs to do in the dephasing lead model).

A recent paper by Sokolov \cite{Sok04} studies a similar geometry, a quantum dot connected to a long lead closed at one end, but in that work there are no voltage fluctuations in the stub. Energy averaging can still suppress certain quantum interference effects (such as the universal conductance fluctuations), but not others (such as weak localization).

\ack
This project grew out of discussions with H. Schomerus. It was supported by the Dutch Science Foundation NWO/FOM.

\end{document}